\documentclass[fleqn,usenatbib]{mnras}
\usepackage{newtxtext,newtxmath}

\usepackage[T1]{fontenc}
\usepackage{ae,aecompl}

\usepackage{graphicx}	
\usepackage{amsmath}	
\usepackage{amssymb}	
\usepackage{subfigure}

\title[High efficiency OH suppression]{First demonstration of OH suppression in a high efficiency near-infrared spectrograph}

\author[Ellis et al.]{S.C.~Ellis$^{1,2}$\thanks{E-mail: simon.ellis@mq.edu.au},  
J.~Bland-Hawthorn$^{2,3}$,  
J.S.~Lawrence$^1$, 
A.J.~Horton$^1$,   
R.~Content$^{1}$,\newauthor
M.M.~Roth$^4$,
N.~Pai$^{1}$,
R.~Zhelem$^1$,
S.~Case$^1$,
E.~Hernandez$^4$,
S.G.~Leon-Saval$^{2,3}$,\newauthor
R.~Haynes$^{5}$,
S.S.~Min$^{3}$,
D.~Giannone$^{4}$,
K.~Madhav$^{4}$,
A.~Rahman$^{4}$,
C.~Betters$^{3}$, \newauthor
D.~Haynes$^{5}$, 
W.~Couch$^{1}$,
L.J.~Kewley$^{5}$,
R.~McDermid$^{6}$,
L.~Spitler$^{6}$,
R.G.~Sharp$^{5}$,\newauthor
and S.~Veilleux$^{7,8}$\\
$^{1}$Australian Astronomical Optics (AAO), Faculty of Science and Engineering, Macquarie University, NSW 2109, Australia\\
$^{2}$Sydney Institute for Astronomy (SIfA), School of Physics A28, University of Sydney, NSW 2006, Australia\\
$^{3}$Sydney Astrophotonic Instrumentation Labs (SAIL), School of Physics A28, University of Sydney, NSW 2006, Australia\\
$^{4}$Leibniz-Institut f\"{u}r Astrophysik Potsdam, An der Sternwarte 16, 14482 Potsdam, Germany\\
$^{5}$Research School of Astronomy and Astrophysics, Australian National University, Weston Creek, ACT 2611, Australia\\
$^{6}$Department of Physics and Astronomy, Macquarie University, Sydney, NSW 2109, Australia\\
$^{7}$Department of Astronomy, University of Maryland, College Park, MD 20742, USA\\
$^{8}$Joint Space-Science Institute, University of Maryland, College Park, MD 20742, USA
}

\date{Accepted XXX. Received YYY; in original form ZZZ}

\pubyear{2019}

\begin{document}
\label{firstpage}
\pagerange{\pageref{firstpage}--\pageref{lastpage}}
\maketitle

\begin{abstract}
Ground-based near-infrared astronomy is severely hampered by the forest of atmospheric emission lines resulting from the rovibrational decay of OH molecules in the upper atmosphere.  The extreme brightness of these lines, as well as their spatial and temporal variability, makes accurate sky subtraction difficult.  Selectively filtering these lines with OH suppression instruments has been a long standing goal for near-infrared spectroscopy.  We have shown previously the efficacy of fibre Bragg gratings combined with photonic lanterns for achieving OH suppression.  Here we report on PRAXIS, a unique near-infrared spectrograph that is optimised for OH suppression with fibre Bragg gratings. We show for the first time that OH suppression (of any kind) is possible with high overall throughput (18 per cent end-to-end), and provide examples of the relative benefits of OH suppression.  
\end{abstract}




\section{Introduction}
\label{sec:intro}

Ground-based near-infrared (NIR) spectroscopy is severely hindered by extremely bright background emission from the earth's atmosphere.  Between 1 and 1.8~$\mu$m, this background is dominated by a forest of emission lines resulting from the rovibrational de-excitation of  OH molecules at $\approx 87$~km altitude (for a review of the NIR background see \citealt{ell08} and references therein).  This background is notoriously difficult to subtract cleanly, since it is very bright ($\approx 14$~mag~arcsec$^{-2}$ in the H band) and highly variable, both temporally and spatially, leading to a large Poissonian noise, and large systematic errors (see e.g.\ \citealt{dav07}).

In 2004, Bland-Hawthorn et al.\ introduced the idea of using fibre Bragg gratings (FBGs) to filter these atmospheric emission lines from observations, and demonstrated the first devices in the laboratory (\citealt{bland04}).  Unlike other OH suppression techniques such as high dispersion masking (e.g.\ \citealt{con94,iwa94,mai00b,iwa01,mot02,par04}), the OH lines are removed \emph{prior} to the light entering the spectrograph, and in a manner dependent solely on wavelength. Subsequent refinements resulted in FBGs capable of suppressing the emission of 150 lines (in most cases actually very closely spaced doublets), at a resolving power of 10,000, by a factor of up to 30~dB over a bandwidth of 400~nm, with both the wavelength and depth of the notches perfectly matched to the OH lines (\citealt{bland08}).  

FBGs only work in the single mode regime, since the different modes of a multimode fibre will not remain in phase, and thus cannot simultaneously constructively interfere.  However, in order to collect enough light at the telescope focus, one must use large core area fibres, which are necessarily multimoded.  Photonic lanterns were developed to solve this problem  (\citealt{leo05,noo09,leo10,noo10,noo12,bir15}), by efficiently converting a multimode fibre into a parallel array of single mode fibres, and vice versa, and thereby enabling FBGs to be incorporated into a spectrograph fibre feed.
The science case and expected performance of OH suppression with FBGs was modelled by \citet{ell08}, and the detailed technique and principles were comprehensively reviewed by \citet{bland11b}.

OH suppression with FBGs was first demonstrated in an on-sky test in which a multimode fibre was pointed directly at the sky, through a hole in the wall of the dome of the Anglo-Australian Telescope (AAT).  This first test used  FBGs suppressing 63 doublets at $R=10,000$.  The output of the multimode fibre was relayed into the IRIS2 spectrograph (\citealt{tin04}), to obtain an $R=2400$ H band spectrum, demonstrating clean suppression of the night sky lines (\citealt{bland09}).  However, the light was gathered with a 60~$\mu$m core fibre feeding a $1\times 7$ photonic lantern, resulting in a large loss due to the mismatch in the number of modes.

Following this, a prototype was developed to integrate FBGs into an instrument fed by a telescope.  The resulting instrument, GNOSIS (\citealt{tri13a}), used an array of $7 \times  50$~$\mu$m core fibres, fed by a microlens array, each of which fed a $1 \times 19$ photonic lantern.  Each SMF of the photonic lanterns was spliced to two FBGs in series, which together suppressed the 103 brightest doublets between $1.47$ and $1.7$~$\mu$m.  The output multimode fibres were arranged into a pseudo-slit, the output of which was re-imaged onto a custom slit mask in the IRIS2 spectrograph.  

The GNOSIS experiment demonstrated the background reduction made possible with photonic OH suppression (see \citealt{ell12a} for full details on the performance of GNOSIS).  The OH lines were suppressed by factors of up to 40~dB for the brightest doublets, and the integrated background between 1.47 and 1.7~$\mu$m was reduced by a factor of 9.  Nevertheless, despite this 
achievement, GNOSIS did not show an improvement in signal-to-noise over conventional NIR spectrographs, nor any reduction in the interline continuum due to reduction of the contamination from scattered OH light.  The lack of signal-to-noise improvement is understood to be the result of retro-fitting the OH suppression unit to an existing spectrograph.  All the optics prior to IRIS2 were warm leading to an increased thermal background, and the overall throughput was low ($\approx 4$ per cent), increasing the relative strength of the detector noise.  Together, these two effects counteracted the improvements made in the night sky background (\citealt{ell12a}).  The lack of reduction in the interline continuum could be due to the presence of real emission of unknown origin (see e.g.\ \citealt{con96,sul12,oli15,ngu16}), which could be dependent on the observing site, but interpretation is confounded by the low signal-to-noise of the measurements.


In order to remedy these deficencies, and thereby fully assess and quantify the performance of OH suppression with FBGs, we have built a new spectrograph and fibre feed, PRAXIS (\citealt{hor12,con14,ell16,ell18}).  PRAXIS is optimised for an OH suppressed fibre feed, and was designed to have low thermal emission and detector noise, whilst maintaining high throughput.  It is fed via a small microlens array feeding 7 OH suppressed fibres (and an additional 12 non-suppressed fibres to assist in object acquisition) which is optimised for measuring the night sky background, whilst also allowing observations of single objects.

PRAXIS has been tested in three commissioning runs. PRAXIS has now demonstrated OH suppression in a high efficiency spectrograph for the first time.  This is an important milestone for OH suppression, since it proves the viability of photonic OH suppression in a practicable instrument, i.e.\  FBGs and photonic lanterns  can be efficiently incorporated into an astronomical instrument, and  there are no inherent impediments in doing so.  

In this paper we provide an interim report describing this significant result.
%
We describe the instrument in section~\ref{sec:praxis}, and report on the performance of the  instrument in section~\ref{sec:perf}, including an analysis of the OH suppression and sky background.  Thereafter we demonstrate the capability and potential of OH suppression through some early science verification observations of Seyfert nuclei
in section~\ref{sec:sv}.  Finally we summarise our results in
section~\ref{sec:conc}.



\section{PRAXIS: instrument description}
\label{sec:praxis}

\begin{figure*}
    \centering
    \includegraphics[scale=0.45]{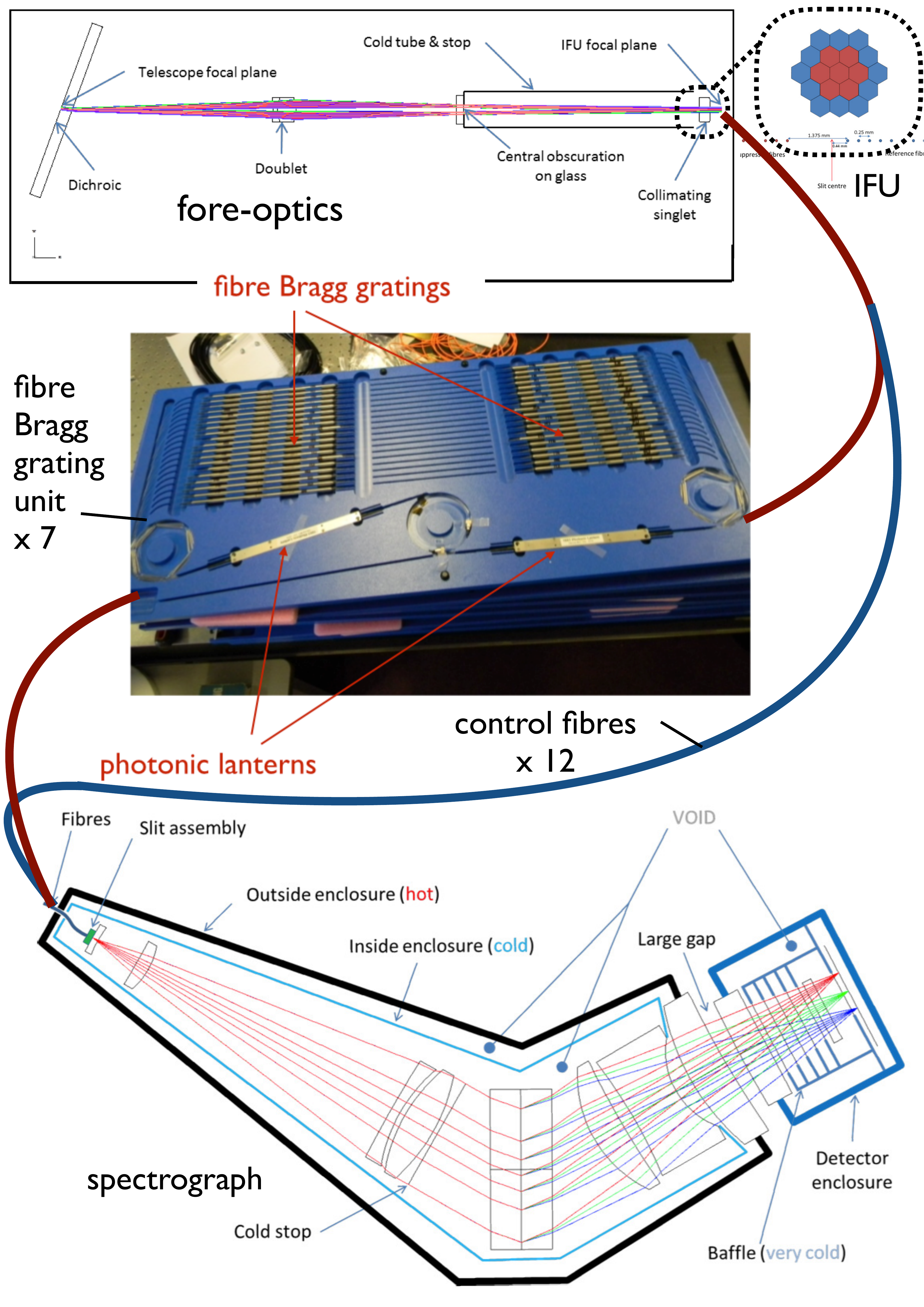}
    \caption{Schematic diagram of PRAXIS.  Light enters from the telescope at the top left.  The main components are the fore-optics (top), the fibre Bragg grating unit (middle) and the spectrograph (bottom).  See the text for a full description.}
    \label{fig:praxis}
\end{figure*}

The PRAXIS instrument design has been described in full in several SPIE papers (\citealt{hor12,con14,ell16,ell18}) and the general scheme of OH suppression was given in the introduction; here we give a brief summary and overview.  The overall scheme of PRAXIS is sketched in Figure~\ref{fig:praxis}.  

PRAXIS has four main components: the fore-optics, the IFU and fibre cable, the fibre Bragg grating unit, and the spectrograph.  We will now describe each of these, beginning with the fibre Bragg grating unit.

\subsection{Fibre Bragg grating unit}

The heart of PRAXIS is the fibre Bragg grating unit, which filters the OH lines from the incoming light, and which sets it apart from all other near-infrared spectrographs.  The design of the rest of the instrument follows from the characteristics of the FBG unit, and so we describe it first.

\begin{figure*}
\centering
\includegraphics[scale=0.78]{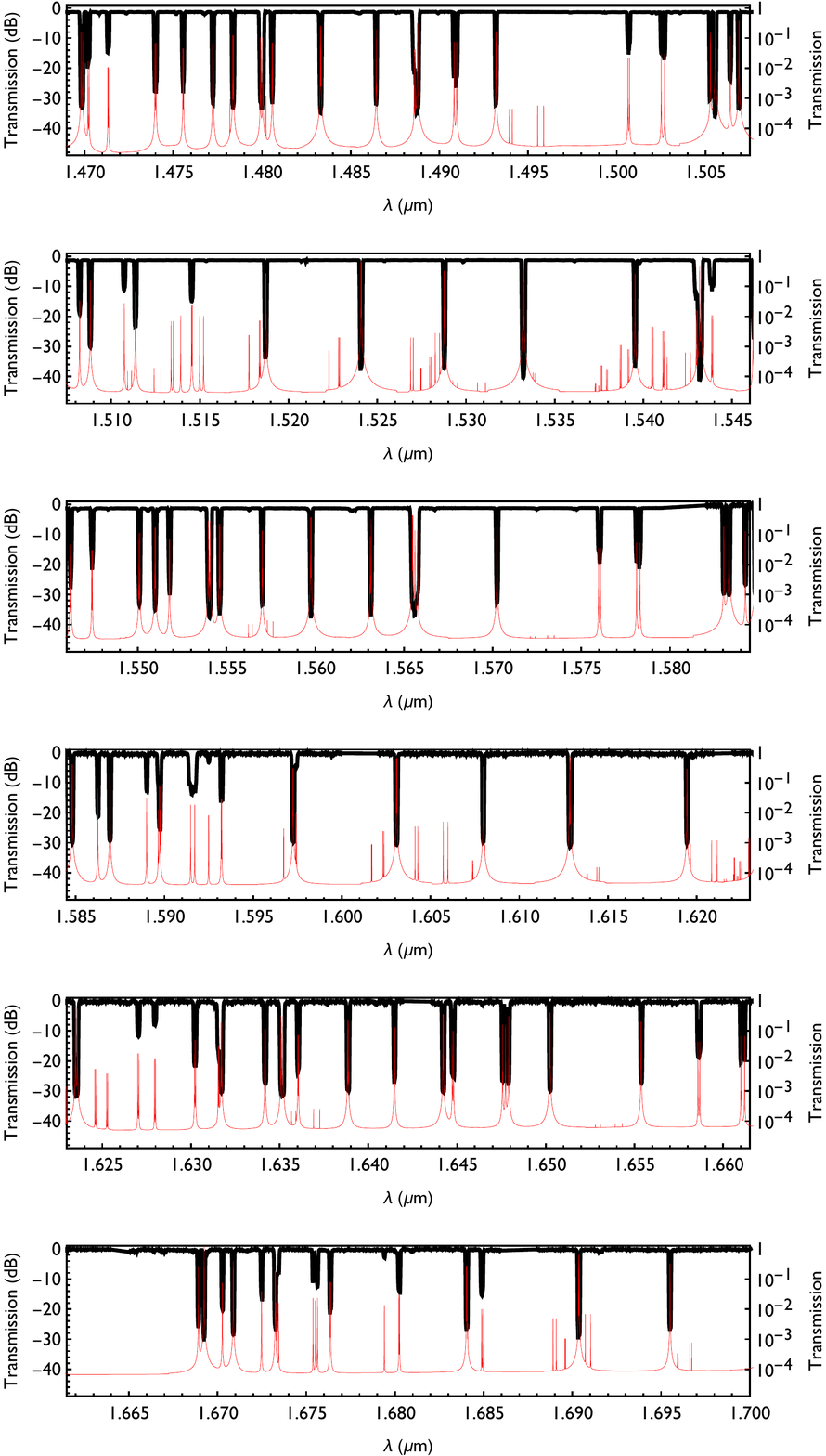}
\caption{The logarithm of the measured transmission of the PRAXIS FBGs (black) compared to the logarithm of a model of the sky surface brightness (red).  The FBG notches match the sky lines with excellent fidelity, in terms of wavelength, depth and width.}
\label{fig:fbgplot}
\end{figure*}

\begin{figure*}
\centering
\includegraphics[scale=0.85]{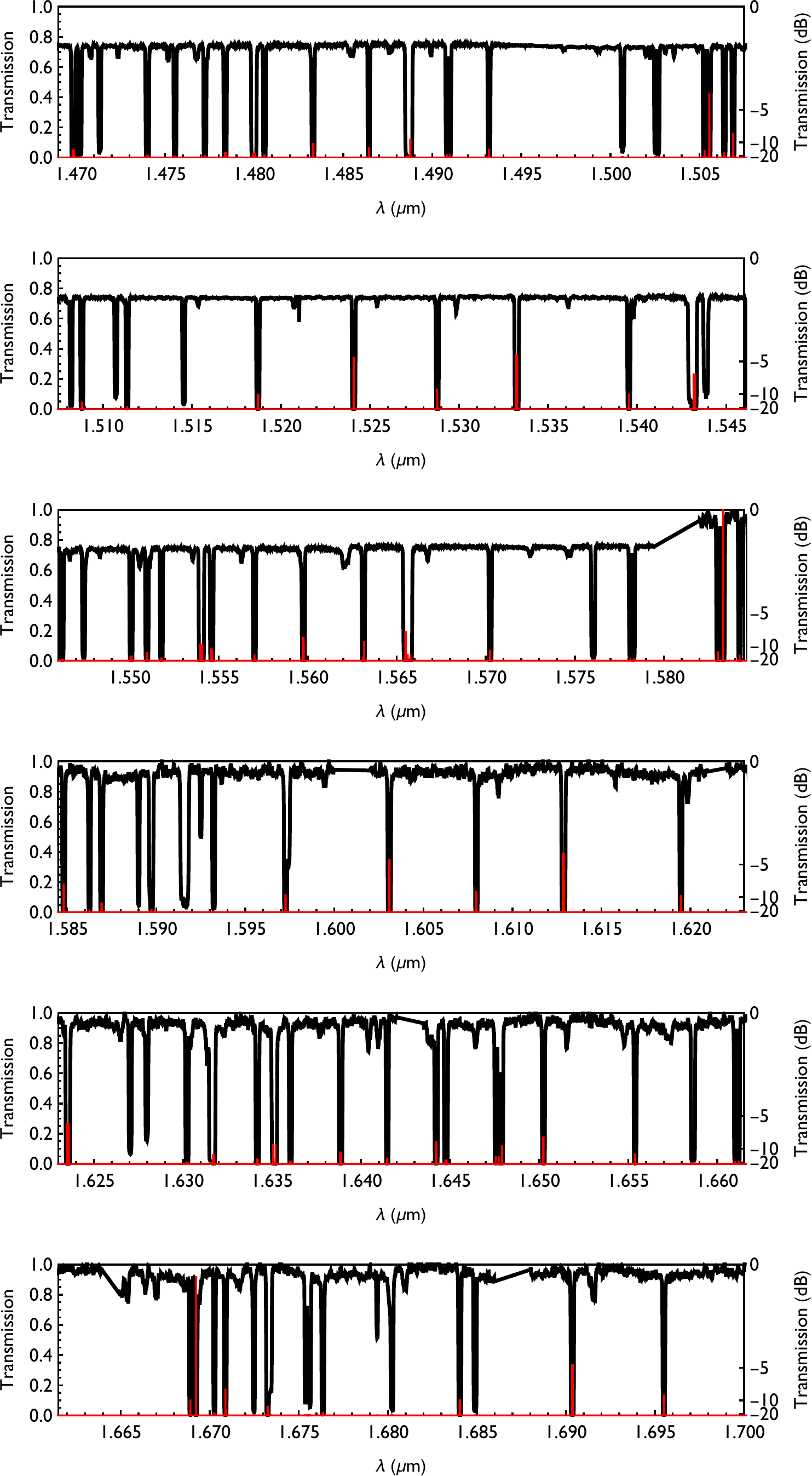}
\caption{As for Fig.~\ref{fig:fbgplot}, but with both the FBG transmission and model sky surface brightness shown on  a linear scale for comparison.}
\label{fig:fbgplotlin}
\end{figure*}

The filtering is done by aperiodic fibre Bragg gratings, the design and principles of which are described by \citet{bland04,bland08,bland11b}.   PRAXIS uses the same FBGs as GNOSIS (\citealt{tri13a}).  Rather than a periodic variation in refractive index in the core of the fibre, which produces a strong reflection at a particular wavelength (and harmonics thereof), the core of the refractive index has an extremely complicated aperiodic variation giving rise to a series of strong reflections at the wavelengths of the OH lines.  PRAXIS uses two aperiodic gratings spliced together to suppress the 103 brightest OH doublets between 1.47 and 1.7~$\mu$m.  The depth of the filter notches is matched to the average strength of the OH lines.  The notches are very square (well fitted by an 8th order Butterworth profile, \citealt{tri13a}), such that notches do not affect the interline regions.  The transmission of the FBGs between the notches is $>90$ per cent, and the typical notch width is $200$~pm, which is approximately two thirds of a pixel.   The measured transmission of the two FBGs in series is shown in Figures~\ref{fig:fbgplot} and \ref{fig:fbgplotlin}, overlaid with the model OH emission spectrum.

%


As described in the introduction, each FBG is inscribed into a single mode fibre, and therefore for efficient coupling to a telescope the FBGs must be incorporated into a photonic lantern.  At the time when the GNOSIS FBGs were being made, the most efficient photonic lanterns had 19 single mode fibres, and therefore these were selected.  In order to increase the total number of modes, and thereby the field-of-view, 7 photonic lanterns were made, each containing identical FBGs.  For efficient performance the multimode input of each photonic lantern must  be fed at the correct focal ratio to excite approximately 19 modes.  In fact the throughput of a photonic lantern is a non-trivial function of the input focal ratio, but has been well characterised empirically (\citealt{hor14}).   Consideration of this throughput dependence is of central importance in the design of the fore-optics, which will be discussed next.

\subsection{Fore-optics}

The primary purpose of the fore-optics is to couple light from the f/15 Cassegrain focus of the AAT  into the multimode inputs of the photonic lanterns.  The injection follows the standard pupil imaging technique of many fibre array integral field units (\citealt{ren02}) in which a microlens array segments the image plane, and each microlens forms a pupil image on the front face of the fibre.  Before the microlens array are some relay optics which change the plate-scale from the default f/15 focus to provide the correct sampling for each 250~$\mu$m wide hexagonal microlens.

For PRAXIS the appropriate sampling for the microlenses is that which maximises the signal-to-noise in the fibres, which must take into account the transmission of the photonic lanterns as a function of input focal ratio, as well as the dependence of the source and background counts on the area of the sky subtended by each fibre.  This optimisation differs for a point source or for a constant surface brightness.  Since one of the primary science cases for PRAXIS is to measure the night sky background with OH suppression, including the interline continuum, the optimal spatial sampling was chosen as 0.55 arcsec per microlens, corresponding to a fibre injection speed of f/4, which is a compromise between optimising for the sky brightness whilst still allowing scientific observations of individual sources.   

To increase the field-of-view, PRAXIS has a  small IFU of 7 fibres, each feeding its own photonic lanterns, so the total field of view is therefore 1.65 arcsec across, or an area of 1.8 arcsec$^{2}$.    In addition, the IFU has an outer ring of 12 fibres which are not OH suppressed, but lead directly to the spectrograph, see Figure~\ref{fig:praxis}.  These fibres provide a larger field-of-view to facilitate acquisition, and are also useful to provide  control spectra for comparison with the OH suppressed spectra.  The 12 non-suppressed fibres are physically separated from the OH suppressed fibres at the slit to avoid cross-talk between adjacent spectra.

The fore-optics unit also contains a dichroic beam-splitter as the first optical element, which reflects light with $\lambda < 1.05$~$\mu$m to a visible acquisition and guiding camera.  The relay-optics contain an intermediate pupil, at which point there is a cold-stop at 262~K to baffle thermal emission from outside the beam, and from the telescope central obstruction.

\subsection{Fibre cable}

The microlens array feeds an array of 19 fibres each with a core diameter of 50~$\mu$m and a numerical aperture (NA) of 0.2.  The fibre cable is 15~m long in order to reach the floor of the AAT dome, where the spectrograph is situated, from the Cassegrain focus.  The fibres are terminated with FC/PC connectors, which connect to the input of the FBG unit for the inner seven fibres, or directly to the spectrograph fibre cable for the outer 12.  The output of the FBG unit also connects to the spectrograph fibre cable with FC/PC connectors.

\begin{figure}
    \centering
    \includegraphics[scale=0.5]{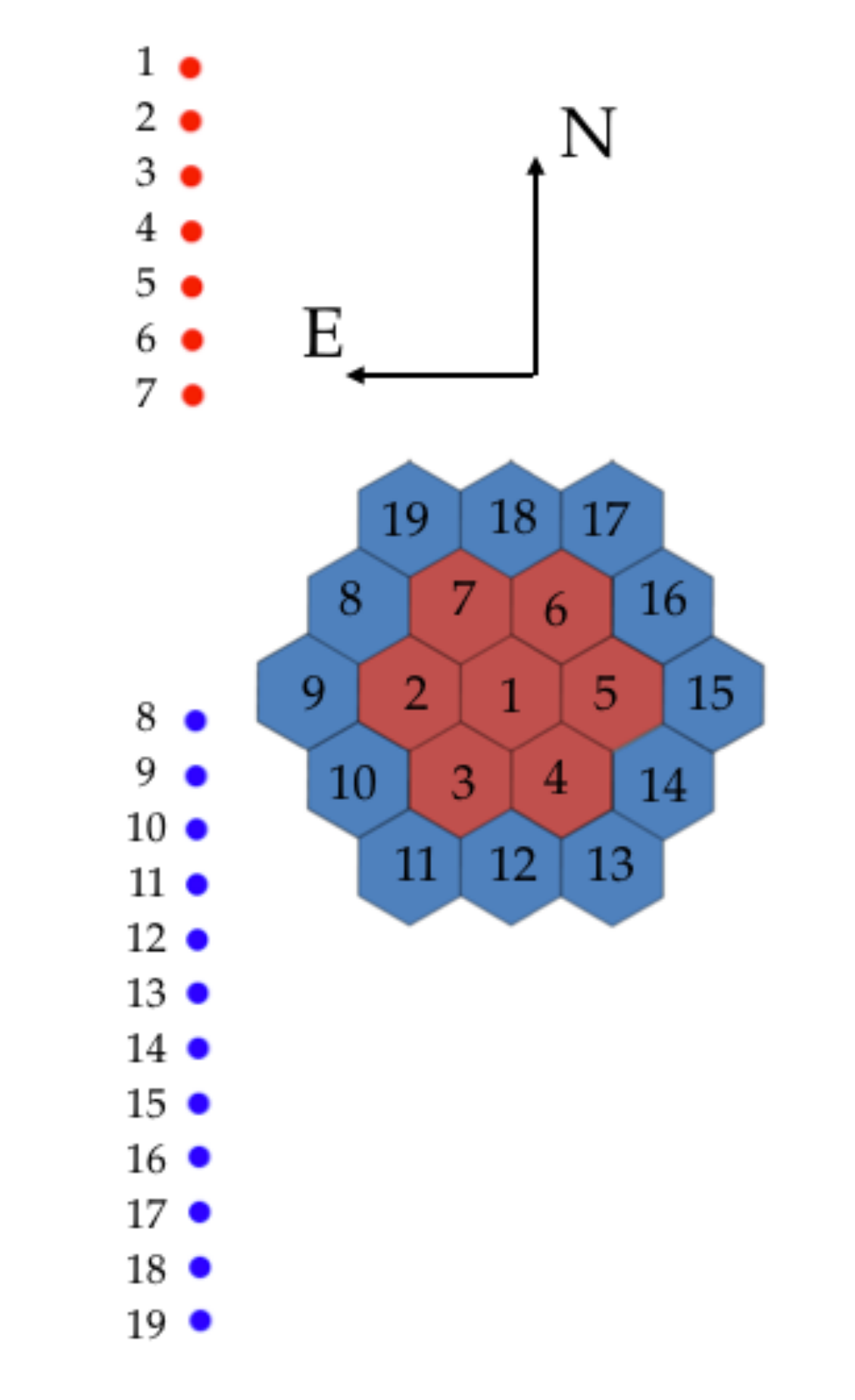}
    \caption{The orientation of the IFU and the mapping of the fibres onto the spectrograph slit.  The non-suppressed fibres are separated from the OH suppressed fibres to avoid cross-contamination.}
    \label{fig:ifumap}
\end{figure}

The fibres enter the spectrograph pre-slit area via a vacuum feed-through.  The fibres are terminated in a slit block, which forms the entrance slit of the spectrograph.  The mapping of the IFU to the slit is shown in Figure~\ref{fig:ifumap}.  Note that the non-suppressed fibres are separated from the suppressed fibres by $30$ times the fibre core diameter to ensure there is no cross-talk between the fibres.

\subsection{Spectrograph}

Much of the improved performance of PRAXIS compared to GNOSIS is due to the spectrograph, which was designed specifically for the  FBG fibre feed, with increased throughput, low detector noise, and a low thermal background.  The increased throughput and low detector noise have been demonstrated, but the low thermal background has not yet been demonstrated on-sky; we describe these measurements in \S\ref{sec:perf}, and here we describe the design.

The PRAXIS spectrograph has a fixed format designed to cover only the wavelengths of the GNOSIS FBGs, i.e.\ 1.47 -- 1.7~$\mu$m, and therefore all components, including glasses, coatings, and the VPH grating were   optimised for this wavelength range.  It uses a state-of-the-art Hawaii-2RG HgCdTe Teledyne detector, housed in its own cryostat, with an ASIC sidecar controller, provided by GL Scientific.  These detectors have high quantum efficiency, low read out noise and low dark current; see \S\ref{sec:det}.

PRAXIS is designed to have low thermal background by cooling all significantly emitting parts.  The spectrograph optics are all enclosed in a radiation shield cooled to 123.5~K with a SunPower cryocooler.   The detector, is housed in a separate  dewar and operated at 75~K, with the radiation shield, including baffles is at 125~K.

\section{PRAXIS: performance}
\label{sec:perf}

The on-sky performance of PRAXIS has been demonstrated in three commissioning runs.  The instrument is not yet fully commissioned, but a number of important milestones for OH suppression have now been demonstrated.  

The  commissioning runs are summarised in Table~\ref{tab:comm}.  The assembly and integration of the instrument was achieved without major difficulty, and all standard observing procedures, including acquisition, guiding, calibrations, and  data acquisition have been established.  Below we describe the performance of PRAXIS, demonstrating that effectual OH suppression is now achievable.

\begin{table*}
    \centering
    \caption{Summary of the first two PRAXIS commissioning runs.}
    \begin{tabular}{lll}
         Dates & Number of clear nights & Achievements  \\ \hline 
         26th -- 31st July 2018 & 2 & Integration on telescope\\
         && Image quality\\
         && Acquisition with offset guide camera\\
         && Offset guiding\\
         && Image reconstruction software \\
         && Data acquistion and detector operation\\
         && Wavelength calibration\\
         && Flat-field calibration \\
         \hline 
          16th -- 21st October 2018 & 2 & Acquisition and guiding with Cassegrain unit camera\\
          && Throughput \\
          && Scientific verification \\
          && Data reduction procedures\\
         \hline 
          15th -- 21st July 2019 & 5 & Measurement and characterisation of thermal emission
    \end{tabular}
    \label{tab:comm}
\end{table*}



\subsection{Throughput}

The end-to-end throughput of PRAXIS has been measured on-sky using photometric standard stars, by which we mean the throughput from the top of the atmosphere to the detector, including the effects of the telescope and the atmosphere as well as PRAXIS itself.  Stars of type A0V were observed, and the data were reduced following standard procedures written in Mathematica.  A data reduction pipeline for PRAXIS is now available in P3D (\citealt{san10}).

The throughput was estimated as follows.   The \citet{cas94} Vega model was scaled to the appropriate H band magnitude of the star observed, and then converted to ADU per pixel by multplying by the exposure time, the collecting area (taking into account the central obstruction of 15 per cent by area for the f/15 top-end of the AAT), the dispersion per pixel, the detector gain, and correcting for the aperture losses.  The throughput of the entire optical train from the atmosphere to the detector is then found by dividing the observed spectrum by the model spectrum.

\begin{figure*}
    \centering
    \subfigure[HIP 101106 -- 18th October, seeing = 1.1 arcsec]{
    \includegraphics[scale=0.9]{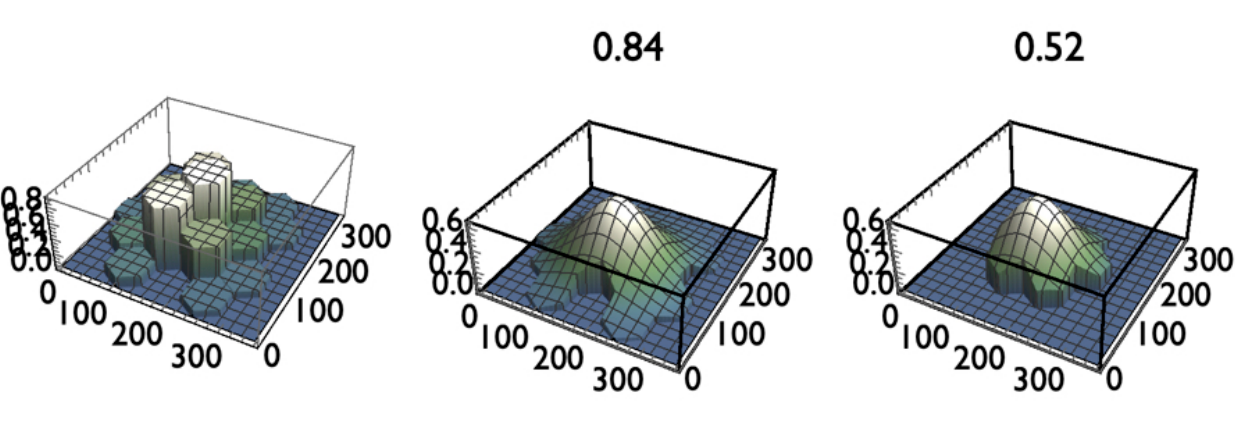}}
    \subfigure[HIP 209960 -- 21st October, seeing = 2.4 arcsec]{
    \includegraphics[scale=0.9]{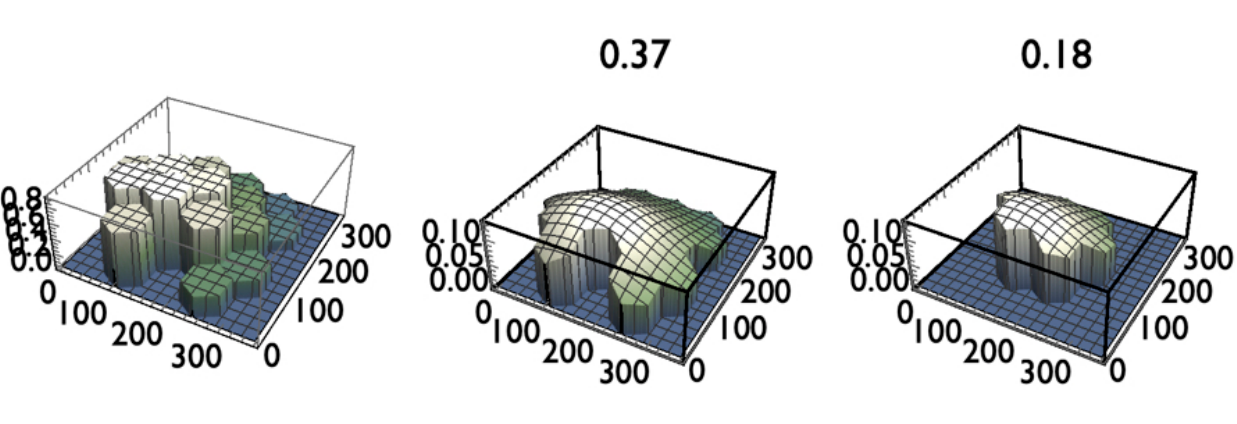}}
    \subfigure[HIP 110963 -- 21st October, seeing = 1.6 arcsec]{
    \includegraphics[scale=0.9]{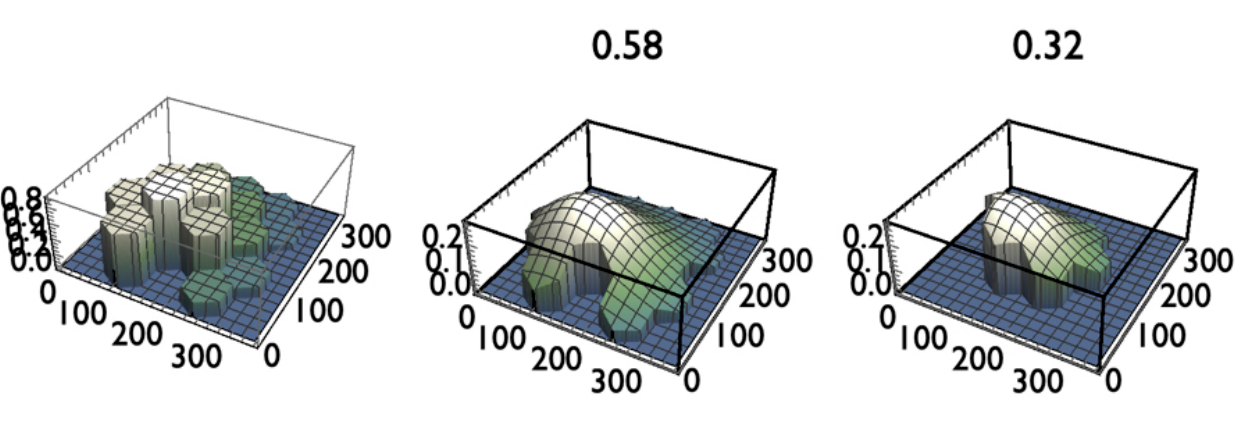}}
    \caption{The reconstructed images of the standard stars (left); the best fitting Moffat profile and total fractional flux enclosed in the IFU (middle); the best fitting Moffat profile and fractional flux enclosed in the centre 7 fibres (right).}
    \label{fig:aploss}
\end{figure*}

The most uncertain step in this calculation is the correction for aperture losses, which depends on both the seeing and the centring of the object during the observation.  This was calculated in the following way.  First, the centroid of the star in the IFU was determined from the centre-of-gravity of the fluxes in each spaxel.  Then a Moffat profile PSF was fit to the fluxes in each spaxel, including the unsuppressed outer 12 fibres, using the best fit centre of gravity and 1 arcsec seeing as initial estimates.  The best fitting PSF was compared to estimates of the seeing made during the night for consistency.  Thereafter the fraction of flux enclosed within the inner 7 OH suppressed fibres was calculated from the best fitting Moffat profile.  The results of this process for HIP~101106, HIP~209960, and HIP~110963 observed on 18th and 21st October are shown in Figure~\ref{fig:aploss}.  The best fitting seeing was 1.1 arcsec on 18th October, and 1.6 -- 2.4 arcsec on 21st October, consistent with estimates made from the acquisition and guiding camera throught the night. 

\begin{figure}
    \centering
    \includegraphics[scale=0.6]{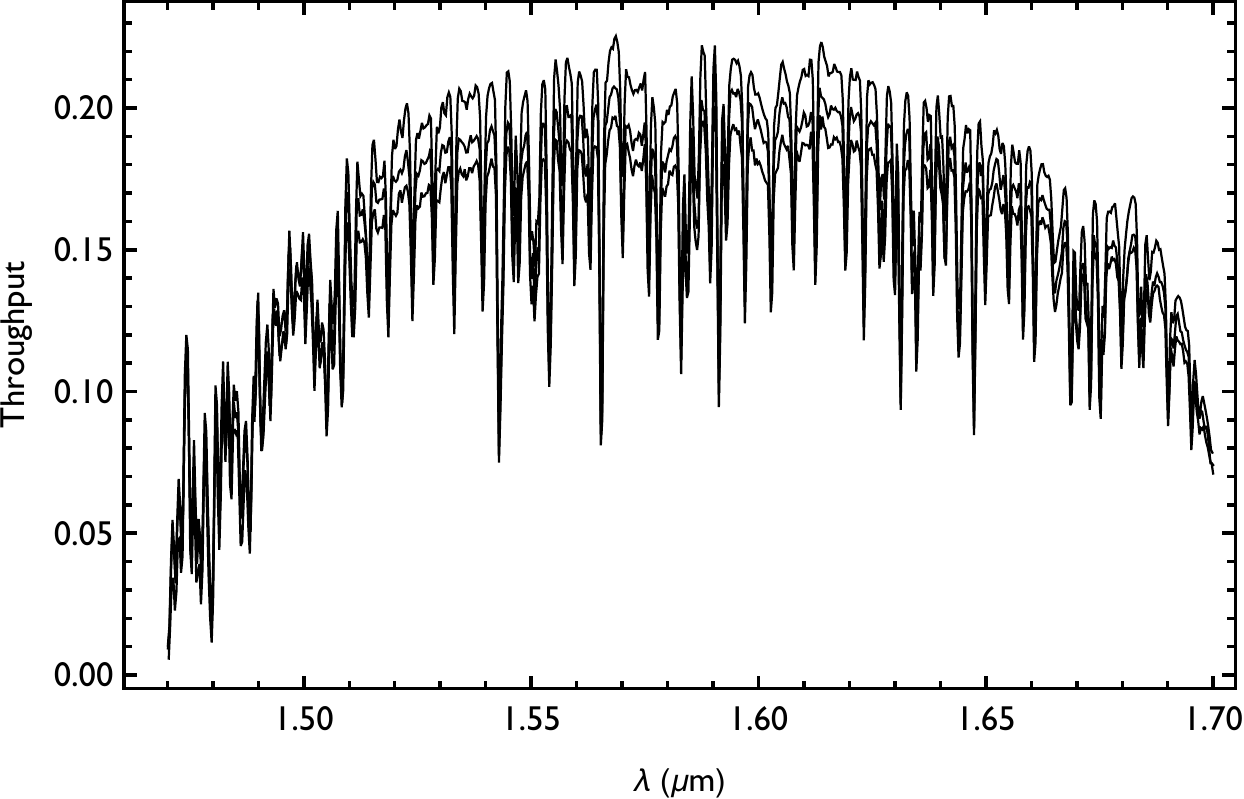}
    \caption{The measured throughput, from the top of the atmosphere to the detector, from the three standard stars in Figure~\ref{fig:aploss}.  The dips are due to the FBG filter notches, convolved with the resolving power of the spectrograph.}
    \label{fig:through}
\end{figure}

The resulting measured throughputs as a function of wavelength are shown in Figure~\ref{fig:through}.  Excluding the FBG notches, the average end-to-end throughput from telescope to detector between 1.55 and 1.65~$\mu$m is $18 \pm 1$ per cent.

\begin{figure}
    \centering
    \includegraphics[scale=0.46]{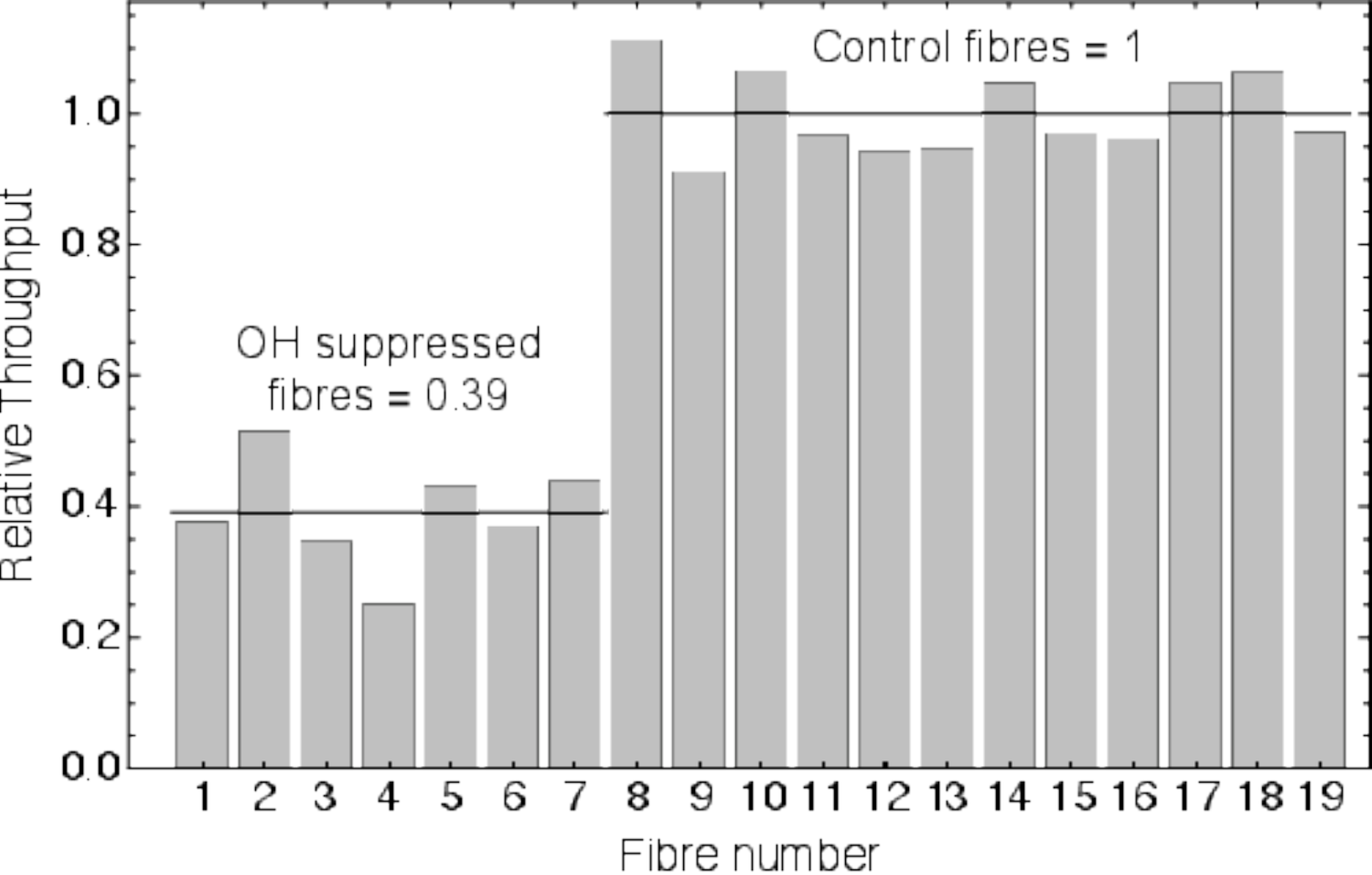}
    \caption{The relative throughput of each fibre from flat-field exposures, excluding the effect of the FBG notches.}
    \label{fig:ffvar}
\end{figure}

The throughput described above includes the effect of the FBG unit which provides the OH suppression.  The FBG unit itself has a throughput of 0.39, which was found by comparing the total counts in each of the suppressed fibres to the total counts in each of the non-suppressed fibres for the fibre flat-field exposures.  For this calculation, the effect of the notches was excluded.  Including the notches the throughput of the FBG unit is 0.36.  Figure~\ref{fig:ffvar} shows a bar chart of the relative throughput of each fibre. 

\subsection{Instrument background}
\label{sec:back}

In order to fully exploit the benefits of OH suppression, PRAXIS was designed to have a high throughput, and a low instrumental background.  As discussed in the previous section, the goal of high efficiency has been achieved.     We now turn our attention to the instrument background.

\subsubsection{Detector background}
\label{sec:det}

PRAXIS uses a state-of-the-art Hawaii-2RG HgCdTe detector from Teledyne, with a 2.5~$\mu$m cut-off.  This is housed in a dedicated dewar provided by GL Scientific, and controlled with an ASIC sidecar controller.  
The performance of the detector was characterised at innoFSPEC AIP prior to commissioning.
The gain, read noise, and dark current were all measured using standard techniques,  for different pre-amp settings.  The best noise characteristics were found with a gain of 2.23~e$^{-}$/ADU.  The read noise was 12.87~e$^{-}$~pix$^{-1}$ for correlated double sampling images, or an effective read noise of $\approx 3.5$~e$^{-}$~pix$^{-1}$ for Fowler sampling with 32 non-destructive reads at the beginning and end of each exposure.  The dark current was measured to be $0.010 \pm 0.003$~e$^{-}$~s$^{-1}$~pix$^{-1}$.

\subsubsection{Thermal background}
\label{sec:thermal}

\begin{figure}
    \centering
    \includegraphics[scale=0.32]{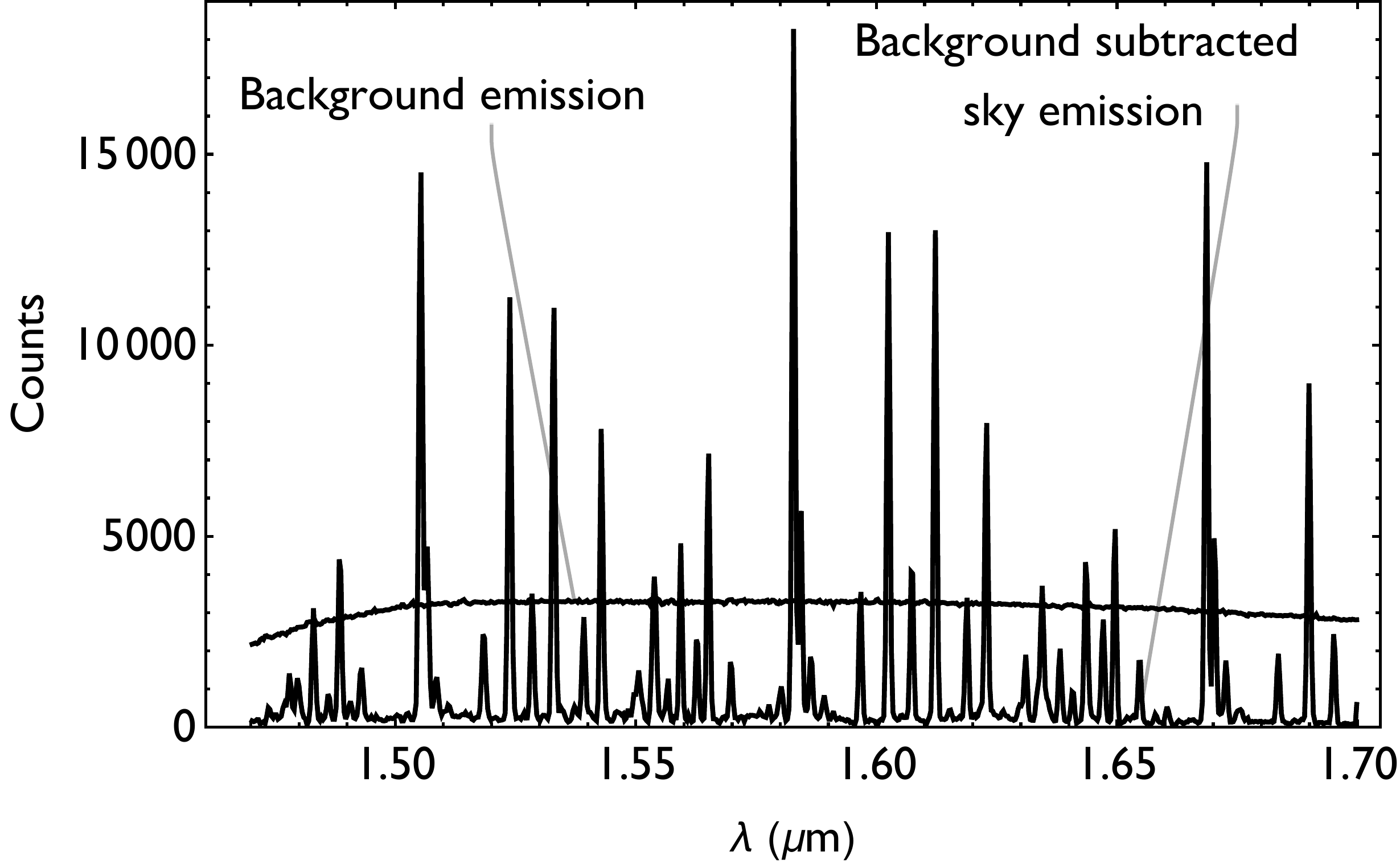}
    \caption{The PRAXIS instrument background near the centre of the detector window, compared to the background subtracted sky.  This excess thermal emission is the limiting factor in the performance of PRAXIS.}
    \label{fig:blob}
\end{figure}

The major limiting factor in the performance of PRAXIS is the thermal background, which is much higher than planned, despite a design intended to reduce this.
Long sky exposures taken during the second and third commissioning runs show that the instrument background is $\approx 16$ times higher than the interline continuum near the centre of the detector window, and $\approx 6$ times higher near the edge of the window, see Figure~\ref{fig:blob}.

Dark exposures taken during the third commissioning run, and during laboratory tests suggest a significant part of this is scattered radiation entering between the windows of the spectrograph and detector, despite the combination of baffles and interference filter designed to block any such thermal emission.  Other tests suggest that some fraction of the emission may also be from the spectrograph itself.  Tests to determine the exact origins of the instrument background are ongoing.  Note that the warm gap was an unavoidable part of the design resulting from the necessity of using a separate turn-key detector system due to practical reasons.

In any case, we wish to emphasise that the radiation is \emph{not} a result of the photonic OH suppression technology.  The background does not originate in the FBGs, nor the photonic lanterns, nor the rest of the fibre cable, but occurs between the fibre traces.  Rather, it is likely that the ultimate cause is either a result of compromising on the design of the spectrograph with warm windows and a gap between the spectrograph camera and the detector dewar, or inadequate baffling inside the spectrograph, or inadequate blocking of long wavelength thermal radiation, or some combination of all these.  With proper baffling and filtering of thermal radiation all these issues can be ameliorated.

\subsection{OH suppression}
\label{sec:ohupp}

\begin{figure*}
    \centering
    \includegraphics{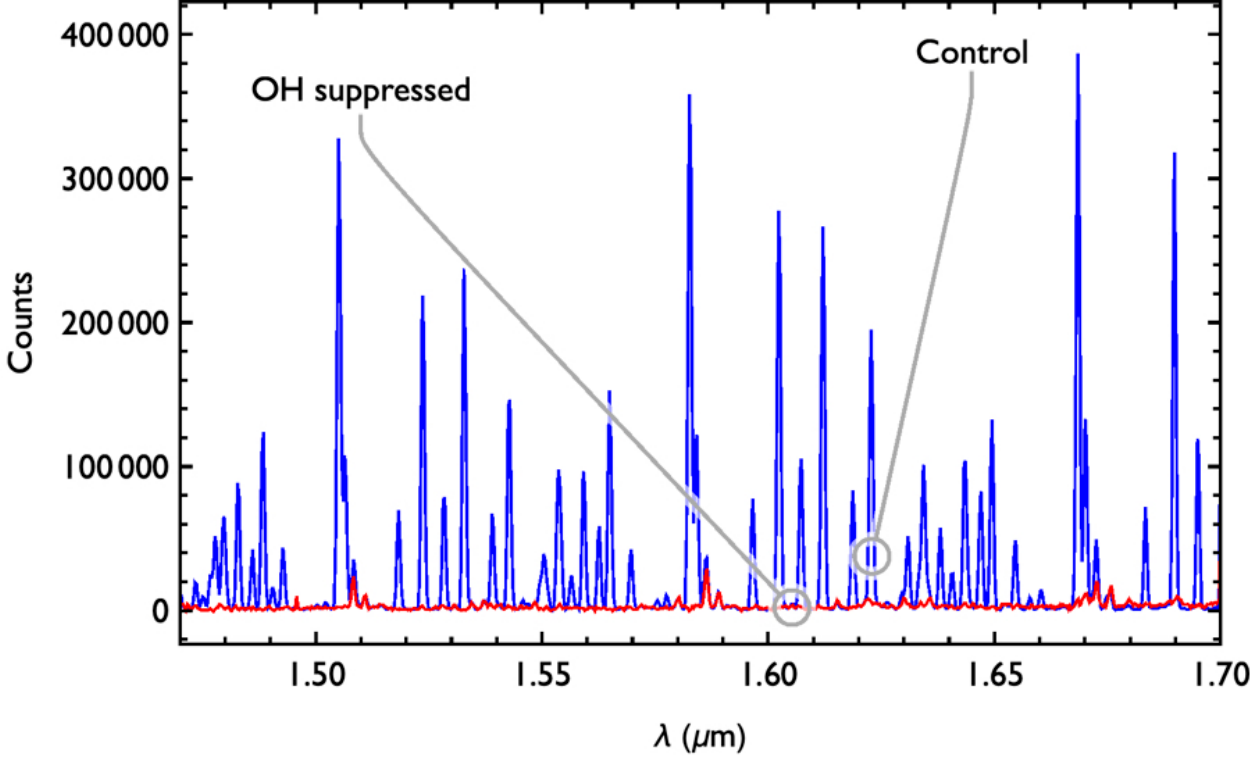}
    \caption{The night sky spectrum measured by PRAXIS, for the OH suppression fibres (red), and the non-suppressed control fibres (blue).}
    \label{fig:ohsupp}
\end{figure*}

The efficacy of OH suppression with fibre Bragg gratings has already been demonstrated with GNOSIS, resulting in a reduction of  integrated background between 1.47 and 1.7~$\mu$m by a factor of 9 (\citealt{ell12a}).  Figure~\ref{fig:ohsupp} shows the night sky spectrum as measured by PRAXIS through the OH suppression fibres (red line) and through the control fibres (blue line).  The integrated background of PRAXIS is a factor of $\sim 8$ lower with the OH suppression fibres, after subtracting off the instrument thermal emission.  However,  there are significant systematic uncertainties in the estimation and subtraction of the instrument thermal background emission, which is quite large.  Therefore, the overall reduction in background should be taken as indicative, and the GNOSIS value of $\approx 9$ should be considered more accurate.  

\section{Science observations}
\label{sec:sv}

The general performance of PRAXIS has been presented in Section~\ref{sec:perf}. 
We now show the potential of OH suppression with  example science observations.   
\emph{It should be borne in mind that the observations presented in this section do not represent the full potential of OH suppression.}  The PRAXIS background is dominated by very strong instrumental thermal emission, and therefore the full benefit of suppressing the sky background is not apparent.  Nevertheless, it is instructive to show the results and process of OH suppression with a science observation.

\subsection{Seyfert galaxies}

There are compelling scientific reasons for obtaining H band spectra of Seyfert galaxies.   Observations are less affected by dust extinction in the NIR than in the visible (\citealt{car89}); the H band contains several diagnostic spectroscopic features, such as  [FeII] emission line at 1.644~$\mu$m and other weaker lines, which indicate photoionisation or shock ionisation (\citealt{sch98}), and provide a useful measurement of the kinematics of the ionised gas; the H band also covers the transition between continuum emission from the central source, which dominates at visible wavelengths, and a stellar and dust dominated thermal continuum (\citealt{mar10}); 
information on the stellar populations can be obtained from the absorption line features e.g.\ CO, Si~I and Mg~I can together 
provide a good classification of spectral type (\citealt{ram06}).

However, in the case of PRAXIS, the main reason Seyfert galaxies were selected to be observed was 
(i) they are relatively bright, and PRAXIS was not performing to full specification due to instrumental thermal emission (\S~\ref{sec:thermal}), (ii) the H band spectra of Seyferts contain several emission and absorption line features providing a useful means to assess the  impact of OH suppression.


We observed the bright Seyfert galaxy, NGC~7674 on 18th October 2018.  The  exposure time was $3 \times 647$~s, with an equal amount of time spent observing sky.

\begin{figure*}
    \centering
    \includegraphics[scale=0.6]{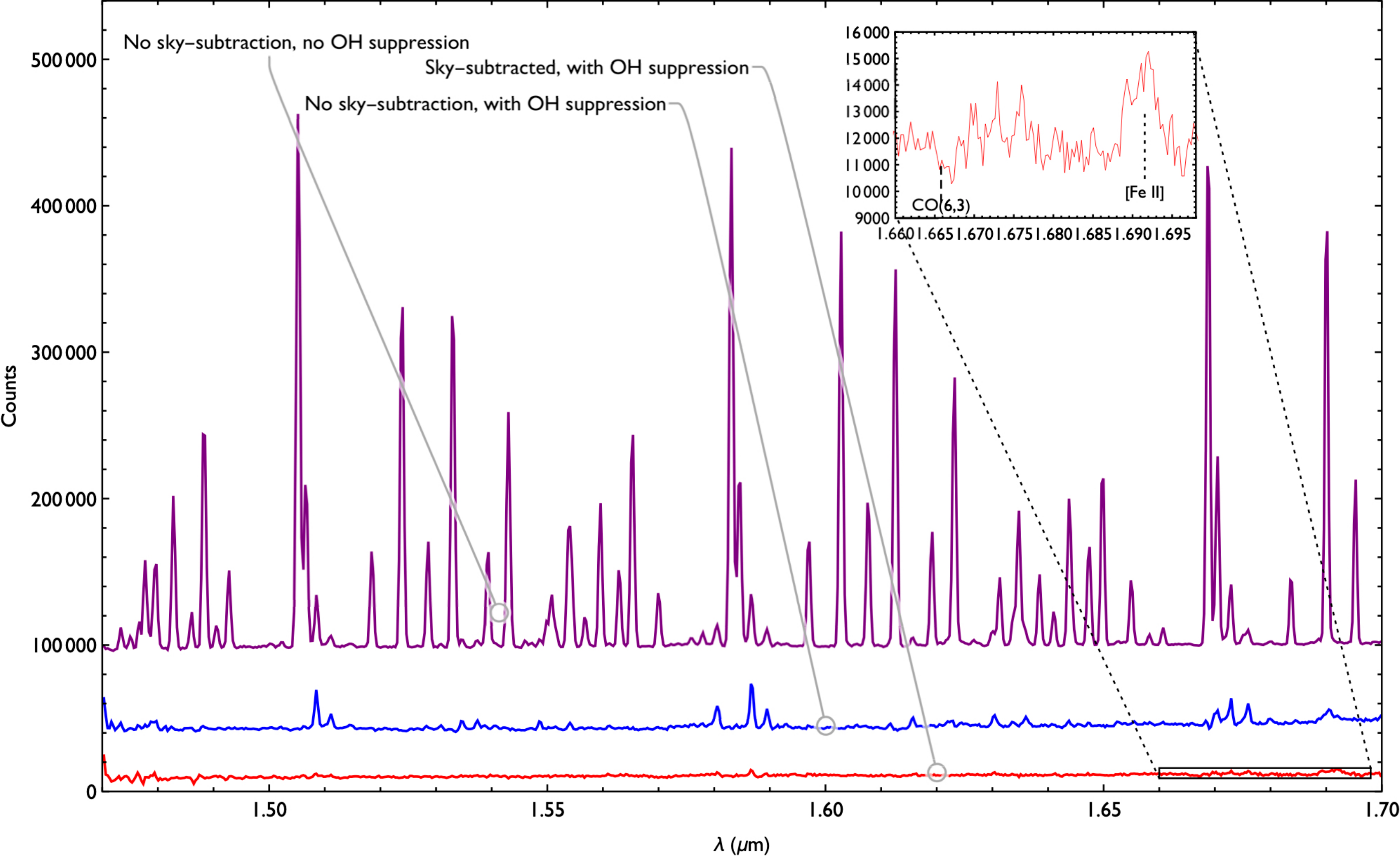}
    \caption{Spectra of NGC~7674 showing different stages of sky-subtraction and OH suppression.  The spectra have been thermal-subtracted, and are offset for clarity.  The inset plot shows the region around the 1.644~$\mu$m [FeII] emission line as an example of the features visible in the final spectrum.}
    \label{fig:ngc7674steps}
\end{figure*}

The spectrum of NGC~7674 is shown in Figure~\ref{fig:ngc7674steps} for different stages of sky-subtraction and OH suppression to illustrate the  advantage of OH suppression.  All spectra have had the thermal background subtracted. 
The comparison of non-sky subtracted spectra  illustrates clearly the benefit of OH suppression; the starting-point for the OH suppressed spectrum (blue) contains far less background than the corresponding control spectrum (purple); the sky-subtracted OH spectrum (red) is therefore much less affected by background noise and sky-residuals.  

The OH suppressed spectra show clear [FeII] emission lines as well as  CO(6,3),  CO(7,3), and Mg absorption lines, see for example the inset plot.  
We note that the exposure times were relatively long for typical H band observations, which is possible since the systematic sky-subtraction errors from varying OH line brightness associated with long exposure times is no longer a problem; this has a considerable operational advantage, since longer exposure times reduce detector read noise and minimise overheads lost in beam-switching.



\section{Conclusions}
\label{sec:conc}

A major success of PRAXIS is the demonstration of OH suppression in a high efficiency spectrograph for the first time.  GNOSIS had already demonstrated clean suppression of the night sky emission lines, but in a low throughput instrument.  This was always understood to be caused by the retro-fitting of the photonic OH suppression technology to an existing near-infrared spectrograph, and moreover one which was not desgined for a fibre feed.  With PRAXIS, we have shown that OH suppression is indeed achievable at high efficiency, and the photonic components have no deleterious effect on performance.

The major limitation of PRAXIS is the high thermal emission.  
This is probably due to inadequate filtering at
long wavelengths, or inadequate shielding
of radiation, either within the spectrograph, or most likely
between the spectrograph and detector dewars.  Due to practical constraints a separate turn-key detector system was a necessary part of the design, and the warm gap was therefore unavoidable. Despite a careful design to mitigate the effects of this gap, extraneous thermal emission was always a risk of this compromise.   Note that we are planning on eliminating this warm interface as the next step of the project.

We wish to reiterate that the thermal emission is not associated with the OH suppression technology.  There is no reason why an FBG OH suppression spectrograph cannot be made with low thermal background.  We have already demonstrated the effectiveness of OH suppression in reducing the sky background (\citealt{ell12a}).  Here we have demonstrated that this is possible in a high efficiency spectrograph.  Re-designing PRAXIS to have a low thermal background will finally enable the true potential of photonic OH suppression to be exploited.

We have not yet been able to assess the effect of OH suppression on the value of the interline continuum.  In \citet{bland04} and \citet{ell08} we  argued that OH suppression should significantly lower the interline continuum, since some large fraction of typical measured values will be due to scattered OH light, some support for which was found in a detailed study of the NIR night sky spectrum by \citet{sul12}.  \citet{oli15}  stated that following the GNOSIS results we retracted this claim in \citet{ell12a}; in fact the thermal emission in both GNOSIS and PRAXIS have prevented an accurate characterisation of the true interline continuum, and this issue is not yet settled.

Finally, we wish to address a common misunderstanding of OH suppression.  It has often been suggested to us in private communication that the FBG  notches will render the filtered parts of the spectrum unusable, and therefore negate any benefit of filtering the emission lines.  However, the OH lines are filtered at high resolution  ($\Delta\lambda \approx 0.18$~nm), which is only $\sim0.5$ of a pixel, and thus the OH lines do not get smeared over several pixels due to the instrument PSF or scattering (\citealt{ell08}).  Moreover, the instrument response is calibrated using fibre flat-field exposures in the normal way.  Examination of the  spectra of NGC~7674 (Figure~\ref{fig:ngc7674steps}) shows that the notches are indeed not visible in the spectra, and that all spectral details are in fact recoverable, unlike the case in which there are significant sky-subtraction residuals.  It remains to demonstrate the same advantage in much fainter objects, which we will do with a future OH suppression spectrograph capitalising on the progress made with GNOSIS and PRAXIS.

\section*{Acknowledgements}

The prototype of the PRAXIS instrument, GNOSIS, was funded under an ARC Federation Fellowship (FF0776384, PI: Bland-Hawthorn) and two ARC LIEF grants LE100100164 and LE120100199. PRAXIS was funded under an ARC Laureate Fellowship (FL140100278, PI: Bland-Hawthorn), and an ARC LIEF grant LE160100191.  innoFSPEC 
acknowledges support from BMBF under grant no. 03Z2AN11.  We thank the SSO Director, Chris Lidman, for allocating Director's time towards PRAXIS commissioning and testing. We thank the staff at the AAT for all their assistance and support during the commissioning of PRAXIS.  We thank the referee for constructive and helpful comments which have improved this paper.





\bibliographystyle{mnras}
\bibliography{ps} 

\begin{thebibliography}{}
\makeatletter
\relax
\def\mn@urlcharsother{\let\do\@makeother \do\$\do\&\do\#\do\^\do\_\do\%\do\~}
\def\mn@doi{\begingroup\mn@urlcharsother \@ifnextchar [ {\mn@doi@}
  {\mn@doi@[]}}
\def\mn@doi@[#1]#2{\def\@tempa{#1}\ifx\@tempa\@empty \href
  {http://dx.doi.org/#2} {doi:#2}\else \href {http://dx.doi.org/#2} {#1}\fi
  \endgroup}
\def\mn@eprint#1#2{\mn@eprint@#1:#2::\@nil}
\def\mn@eprint@arXiv#1{\href {http://arxiv.org/abs/#1} {{\tt arXiv:#1}}}
\def\mn@eprint@dblp#1{\href {http://dblp.uni-trier.de/rec/bibtex/#1.xml}
  {dblp:#1}}
\def\mn@eprint@#1:#2:#3:#4\@nil{\def\@tempa {#1}\def\@tempb {#2}\def\@tempc
  {#3}\ifx \@tempc \@empty \let \@tempc \@tempb \let \@tempb \@tempa \fi \ifx
  \@tempb \@empty \def\@tempb {arXiv}\fi \@ifundefined
  {mn@eprint@\@tempb}{\@tempb:\@tempc}{\expandafter \expandafter \csname
  mn@eprint@\@tempb\endcsname \expandafter{\@tempc}}}

\bibitem[\protect\citeauthoryear{{Birks}, {Gris-S{\'a}nchez}, {Yerolatsitis},
  {Leon-Saval}  \& {Thomson}}{{Birks} et~al.}{2015}]{bir15}
{Birks} T.~A.,  {Gris-S{\'a}nchez} I.,  {Yerolatsitis} S.,  {Leon-Saval} S.~G.,
    {Thomson} R.~R.,  2015, {Advances in Optics and Photonics}, \href
  {http://adsabs.harvard.edu/abs/2015arXiv150302837B} {In press}

\bibitem[\protect\citeauthoryear{Bland-Hawthorn, Englund  \&
  Edvell}{Bland-Hawthorn et~al.}{2004}]{bland04}
Bland-Hawthorn J.,  Englund M.,   Edvell G.,  2004, Optics Express, 12, 5902

\bibitem[\protect\citeauthoryear{{Bland-Hawthorn}, {Buryak}  \&
  {Kolossovski}}{{Bland-Hawthorn} et~al.}{2008}]{bland08}
{Bland-Hawthorn} J.,  {Buryak} A.,   {Kolossovski} K.,  2008, Journal of the
  Optical Society of America A, 25, 153

\bibitem[\protect\citeauthoryear{{Bland-Hawthorn}, {Ellis}, {Haynes}  \&
  {Horton}}{{Bland-Hawthorn} et~al.}{2009}]{bland09}
{Bland-Hawthorn} J.,  {Ellis} S.,  {Haynes} R.,   {Horton} A.,  2009,
  Anglo-Australian Observatory Newsletter, \href
  {http://ads.nao.ac.jp/abs/2006AAONw.110...18E} {115, 15}

\bibitem[\protect\citeauthoryear{{Bland-Hawthorn} et~al.,}{{Bland-Hawthorn}
  et~al.}{2011}]{bland11b}
{Bland-Hawthorn} J.,  et~al., 2011, \mn@doi [Nature Communications]
  {10.1038/ncomms1584}, \href
  {http://adsabs.harvard.edu/abs/2011NatCo...2E.581B} {2, 581}

\bibitem[\protect\citeauthoryear{{Cardelli}, {Clayton}  \& {Mathis}}{{Cardelli}
  et~al.}{1989}]{car89}
{Cardelli} J.~A.,  {Clayton} G.~C.,   {Mathis} J.~S.,  1989, \mn@doi
  [Astrophys.\ J.] {10.1086/167900}, \href
  {http://adsabs.harvard.edu/abs/1989ApJ...345..245C} {345, 245}

\bibitem[\protect\citeauthoryear{{Castelli} \& {Kurucz}}{{Castelli} \&
  {Kurucz}}{1994}]{cas94}
{Castelli} F.,  {Kurucz} R.~L.,  1994, A\&A, \href
  {https://ui.adsabs.harvard.edu/abs/1994A&A...281..817C} {281, 817}

\bibitem[\protect\citeauthoryear{{Content}}{{Content}}{1996}]{con96}
{Content} R.,  1996, Astrophys.\ J., 464, 412

\bibitem[\protect\citeauthoryear{{Content} \& {Angel}}{{Content} \&
  {Angel}}{1994}]{con94}
{Content} R.,  {Angel} J. R.~P.,  1994, ] {10.1117/12.176730}, \href
  {https://ui.adsabs.harvard.edu/abs/1994SPIE.2198..757C} {2198, 757}

\bibitem[\protect\citeauthoryear{{Content} et~al.,}{{Content}
  et~al.}{2014}]{con14}
{Content} R.,  et~al., 2014, Proc.\ SPIE, 9151, 4

\bibitem[\protect\citeauthoryear{{Davies}}{{Davies}}{2007}]{dav07}
{Davies} R.~I.,  2007, MNRAS, 375, 1099

\bibitem[\protect\citeauthoryear{Ellis \& Bland-Hawthorn}{Ellis \&
  Bland-Hawthorn}{2008}]{ell08}
Ellis S.~C.,  Bland-Hawthorn J.,  2008, MNRAS, 386, 47

\bibitem[\protect\citeauthoryear{{Ellis} et~al.,}{{Ellis}
  et~al.}{2012}]{ell12a}
{Ellis} S.~C.,  et~al., 2012, \mn@doi [MNRAS]
  {10.1111/j.1365-2966.2012.21602.x}, \href
  {http://adsabs.harvard.edu/abs/2012MNRAS.425.1682E} {425, 1682}

\bibitem[\protect\citeauthoryear{{Ellis} et~al.,}{{Ellis} et~al.}{2016}]{ell16}
{Ellis} S.~C.,  et~al., 2016, Proc.\ SPIE, 9908, 99084A

\bibitem[\protect\citeauthoryear{{Ellis} et~al.,}{{Ellis} et~al.}{2018}]{ell18}
{Ellis} S.~C.,  et~al., 2018, Proc.\ SPIE, 10702, 107020P

\bibitem[\protect\citeauthoryear{{Horton}, {Ellis}, {Lawrence}  \&
  {Bland-Hawthorn}}{{Horton} et~al.}{2012}]{hor12}
{Horton} A.,  {Ellis} S.,  {Lawrence} J.,   {Bland-Hawthorn} J.,  2012, Proc.\
  SPIE, 8450, 84501V

\bibitem[\protect\citeauthoryear{{Horton}, {Content}, {Ellis}  \&
  {Lawrence}}{{Horton} et~al.}{2014}]{hor14}
{Horton} A.,  {Content} R.,  {Ellis} S.,   {Lawrence} J.,  2014, in Society of
  Photo-Optical Instrumentation Engineers (SPIE) Conference Series. p.~22
  (\mn@eprint {arXiv} {1407.4191}), \mn@doi{10.1117/12.2054570}

\bibitem[\protect\citeauthoryear{{Iwamuro}, {Maihara}, {Oya}, {Tsukamoto},
  {Hall}, {Cowie}, {Tokunaga}  \& {Pickles}}{{Iwamuro} et~al.}{1994}]{iwa94}
{Iwamuro} F.,  {Maihara} T.,  {Oya} S.,  {Tsukamoto} H.,  {Hall} D.~N.~B.,
  {Cowie} L.~L.,  {Tokunaga} A.~T.,   {Pickles} A.~J.,  1994, PASJ, 46, 515

\bibitem[\protect\citeauthoryear{{Iwamuro}, {Motohara}, {Maihara}, {Hata}  \&
  {Harashima}}{{Iwamuro} et~al.}{2001}]{iwa01}
{Iwamuro} F.,  {Motohara} K.,  {Maihara} T.,  {Hata} R.,   {Harashima} T.,
  2001, PASJ, 53, 355

\bibitem[\protect\citeauthoryear{Leon-Saval, Birks, Bland-Hawthorn  \&
  Englund}{Leon-Saval et~al.}{2005}]{leo05}
Leon-Saval S.,  Birks T.,  Bland-Hawthorn J.,   Englund M.,  2005, Optics
  Letters, 30, 19

\bibitem[\protect\citeauthoryear{{Leon-Saval}, {Argyros}  \& {Bland
  -Hawthorn}}{{Leon-Saval} et~al.}{2010}]{leo10}
{Leon-Saval} S.~G.,  {Argyros} A.,   {Bland -Hawthorn} J.,  2010, \mn@doi
  [Optics Express] {10.1364/OE.18.008430}, \href
  {https://ui.adsabs.harvard.edu/abs/2010OExpr..18.8430L} {18, 8430}

\bibitem[\protect\citeauthoryear{{Maihara} et~al.,}{{Maihara}
  et~al.}{2000}]{mai00b}
{Maihara} T.,  et~al., 2000, in {Iye} M.,  {Moorwood} A.~F.,  eds,  Presented
  at the Society of Photo-Optical Instrumentation Engineers (SPIE) Conference
  Vol. 4008, Proc. SPIE Vol. 4008, p. 1111-1118, Optical and IR Telescope
  Instrumentation and Detectors, Masanori Iye; Alan F. Moorwood; Eds.. p.~1111

\bibitem[\protect\citeauthoryear{{Martins}, {Rodr{\'\i}guez-Ardila}, {de Souza}
   \& {Gruenwald}}{{Martins} et~al.}{2010}]{mar10}
{Martins} L.~P.,  {Rodr{\'\i}guez-Ardila} A.,  {de Souza} R.,   {Gruenwald} R.,
   2010, \mn@doi [MNRAS] {10.1111/j.1365-2966.2010.17042.x}, \href
  {https://ui.adsabs.harvard.edu/#abs/2010MNRAS.406.2168M} {406, 2168}

\bibitem[\protect\citeauthoryear{{Motohara} et~al.,}{{Motohara}
  et~al.}{2002}]{mot02}
{Motohara} K.,  et~al., 2002, PASJ, 54, 315

\bibitem[\protect\citeauthoryear{{Nguyen}, {Zemcov}, {Battle}, {Bock},
  {Hristov}, {Korngut}  \& {Meek}}{{Nguyen} et~al.}{2016}]{ngu16}
{Nguyen} H.~T.,  {Zemcov} M.,  {Battle} J.,  {Bock} J.~J.,  {Hristov} V.,
  {Korngut} P.,   {Meek} A.,  2016, \mn@doi [PASP]
  {10.1088/1538-3873/128/967/094504}, \href
  {http://adsabs.harvard.edu/abs/2016PASP..128i4504N} {128, 094504}

\bibitem[\protect\citeauthoryear{{Noordegraaf}, {Skovgaard}, {Nielsen}  \&
  {Bland-Hawthorn}}{{Noordegraaf} et~al.}{2009}]{noo09}
{Noordegraaf} D.,  {Skovgaard} P.~M.,  {Nielsen} M.~D.,   {Bland-Hawthorn} J.,
  2009, \mn@doi [Optics Express] {10.1364/OE.17.001988}, \href
  {http://adsabs.harvard.edu/abs/2009OExpr..17.1988N} {17, 1988}

\bibitem[\protect\citeauthoryear{{Noordegraaf}, {Skovgaard}, {Maack},
  {Bland-Hawthorn}, {Haynes}  \& {L{\ae}gsgaard}}{{Noordegraaf}
  et~al.}{2010}]{noo10}
{Noordegraaf} D.,  {Skovgaard} P.~M.~W.,  {Maack} M.~D.,  {Bland-Hawthorn} J.,
  {Haynes} R.,   {L{\ae}gsgaard} J.,  2010, \mn@doi [Optics Express]
  {10.1364/OE.18.004673}, \href
  {https://ui.adsabs.harvard.edu/#abs/2010OExpr..18.4673N} {18, 4673}

\bibitem[\protect\citeauthoryear{{Noordegraaf}, {Skovgaard}, {Sandberg},
  {Maack}, {Bland-Hawthorn}, {Lawrence}  \& {L{\ae}gsgaard}}{{Noordegraaf}
  et~al.}{2012}]{noo12}
{Noordegraaf} D.,  {Skovgaard} P.~M.~W.,  {Sandberg} R.~H.,  {Maack} M.~D.,
  {Bland-Hawthorn} J.,  {Lawrence} J.~S.,   {L{\ae}gsgaard} J.,  2012, \mn@doi
  [Optics Letters] {10.1364/OL.37.000452}, \href
  {https://ui.adsabs.harvard.edu/#abs/2012OptL...37..452N} {37, 452}

\bibitem[\protect\citeauthoryear{{Oliva} et~al.,}{{Oliva} et~al.}{2015}]{oli15}
{Oliva} E.,  et~al., 2015, \mn@doi [A\&A] {10.1051/0004-6361/201526291}, \href
  {http://adsabs.harvard.edu/abs/2015A%26A...581A..47O} {581, A47}

\bibitem[\protect\citeauthoryear{{Parry} et~al.,}{{Parry} et~al.}{2004}]{par04}
{Parry} I.,  et~al., 2004, Proc.\ SPIE", 5492, 1135

\bibitem[\protect\citeauthoryear{{Ramos Almeida}, {P{\'e}rez Garc{\'\i}a},
  {Acosta- Pulido}, {Rodr{\'\i}guez Espinosa}, {Barrena}  \& {Manchado}}{{Ramos
  Almeida} et~al.}{2006}]{ram06}
{Ramos Almeida} C.,  {P{\'e}rez Garc{\'\i}a} A.~M.,  {Acosta- Pulido} J.~A.,
  {Rodr{\'\i}guez Espinosa} J.~M.,  {Barrena} R.,   {Manchado} A.,  2006,
  \mn@doi [Astrophys.\ J.] {10.1086/504284}, \href
  {https://ui.adsabs.harvard.edu/#abs/2006ApJ...645..148R} {645, 148}

\bibitem[\protect\citeauthoryear{{Ren} \& {Allington-Smith}}{{Ren} \&
  {Allington-Smith}}{2002}]{ren02}
{Ren} D.,  {Allington-Smith} J.,  2002, \mn@doi [PASP] {10.1086/341710}, \href
  {http://adsabs.harvard.edu/abs/2002PASP..114..866R} {114, 866}

\bibitem[\protect\citeauthoryear{{Sandin}, {Becker}, {Roth}, {Gerssen},
  {Monreal-Ibero}, {B{\"o}hm}  \& {Weilbacher}}{{Sandin} et~al.}{2010}]{san10}
{Sandin} C.,  {Becker} T.,  {Roth} M.~M.,  {Gerssen} J.,  {Monreal-Ibero} A.,
  {B{\"o}hm} P.,   {Weilbacher} P.,  2010, \mn@doi [A\&A]
  {10.1051/0004-6361/201014022}, \href
  {https://ui.adsabs.harvard.edu/abs/2010A&A...515A..35S} {515, A35}

\bibitem[\protect\citeauthoryear{{Schreiber}}{{Schreiber}}{1998}]{sch98}
{Schreiber} N.~M.,  1998, PhD thesis, "Ludwig-Maximilians-Universit\"{a}t
  M\"{u}nchen"

\bibitem[\protect\citeauthoryear{{Sullivan} \& {Simcoe}}{{Sullivan} \&
  {Simcoe}}{2012}]{sul12}
{Sullivan} P.~W.,  {Simcoe} R.~A.,  2012, \mn@doi [PASP] {10.1086/668849},
  \href {http://adsabs.harvard.edu/abs/2012PASP..124.1336S} {124, 1336}

\bibitem[\protect\citeauthoryear{{Tinney} et~al.,}{{Tinney}
  et~al.}{2004}]{tin04}
{Tinney} C.~G.,  et~al., 2004, in {Moorwood} A.~F.~M.,  {Iye} M.,  eds,
  Society of Photo-Optical Instrumentation Engineers (SPIE) Conference Series
  Vol. 5492, Ground-based Instrumentation for Astronomy. pp 998--1009,
  \mn@doi{10.1117/12.550980}

\bibitem[\protect\citeauthoryear{{Trinh} et~al.,}{{Trinh}
  et~al.}{2013}]{tri13a}
{Trinh} C.~Q.,  et~al., 2013, \mn@doi [AJ] {10.1088/0004-6256/145/2/51}, \href
  {http://adsabs.harvard.edu/abs/2013AJ....145...51T} {145, 51}

\makeatother
\end{thebibliography}




\bsp	
\label{lastpage}
\end{document}